\documentclass[journal]{IEEEtran}
\pdfoutput=1
\usepackage{amssymb}
\usepackage{subfig}
\usepackage{hyperref}
\usepackage{cite}

\ifCLASSINFOpdf
  \usepackage[pdftex]{graphicx}
  \graphicspath{{../pdf/}{../jpeg/}}
  \DeclareGraphicsExtensions{.pdf,.jpeg,.png}
\else
  \usepackage[dvips]{graphicx}
  \graphicspath{{../eps/}}
  \DeclareGraphicsExtensions{.eps}
\fi
\begin{document}
%
\title{News consumption \\and social media regulations policy}
%
%
%

\author{Gabriele~Etta,
        Matteo~Cinelli,
        Alessandro~Galeazzi,
        Carlo~Michele~Valensise,
        Walter Quattrociocchi,
        Mauro~Conti, \IEEEmembership{Senior Member,~IEEE}

\thanks{}}

\maketitle

\begin{abstract}
Users online tend to consume information adhering to their system of beliefs and to ignore dissenting information.
During the COVID-19 pandemic, users get exposed to a massive amount of information about a new topic having a high level of uncertainty. 
In this paper, we analyze two social media that enforced opposite moderation methods, Twitter and Gab, to assess the interplay between news consumption and content regulation concerning COVID-19. 
We compare the two platforms on about three million pieces of content analyzing user interaction with respect to news articles. 
We first describe users' consumption patterns on the two platforms focusing on the political leaning of news outlets. 
Finally, we characterize the echo chamber effect by modeling the dynamics of users' interaction networks. 
Our results show that the presence of moderation pursued by Twitter produces a significant reduction of questionable content, with a consequent affiliation towards reliable sources in terms of engagement and comments. Conversely, the lack of clear regulation on Gab results in the tendency of the user to engage with both types of content, showing a slight preference for the questionable ones which may account for a dissing/endorsement behavior. Twitter users show segregation towards reliable content with a uniform narrative. Gab, instead, offers a more heterogeneous structure where users, independently of their leaning, follow people who are slightly polarized towards questionable news.
\end{abstract}


%
\IEEEpeerreviewmaketitle

\section{Introduction}
\label{sec:introduction}

The COVID-19 outbreak \cite{who_covid_definition}, which was declared as a pandemic by the World Health Organization (WHO) on 11 March 2020 \cite{who_covid_pandemic}, changed several aspects of our everyday life both in the online and offline sphere. For instance, 
the news diet of users was remarkably modified in its structure by introducing a considerable amount of information referring to a new topic. This phenomenon was accelerated by social media platforms, which are known for shaping discussions on a wide range of issues, including politics, climate change, economics, migration, and health \cite{bessi2015trend,chou2018addressing,bovet2019influence,del2017news}.

The arising of the pandemic generated an overabundant flow of information and news, whose trustworthiness may not always be guaranteed, especially online. This phenomenon, referred as infodemic \cite{zarocostas2020fight,who_infodemic} reportedly affect people's behavior \cite{Sharot2020} in a harmful way. This aspect calls for urgent investigations of the turbulent dynamics of the online infosphere, complementary to the monitoring of the spreading of infections \cite{louis2019,Viboud2802,cinelli2020infodemic}. Indeed, the current infodemic may foster the tendency of users a) to acquire information adhering to their system of beliefs \cite{bessi2015science}, b) to ignore dissenting information \cite{zollo2017debunking}, c) to form polarized groups around a shared narrative \cite{del2016echo}. Two common factors to such behaviors carried on by users are opinion polarization \cite{vicario2019polarization}, one of the dominating traits of online social dynamics, and echo chambers \cite{cinelli2021echochambers}. Divided into echo chambers, users account for the coherence with their preferred narrative rather than the actual value of the information \cite{cinelli2019misinfo, delvicario2016, conti2017s}.
Such evidence for polarization and online echo chambers seems to be related to a feedback loop between individual choices and algorithm recommendations towards like-minded contents \cite{bakshy2015exposure, cinelli2020selective, cinelli2021echochambers}.
However, other presumably harmless factors like the enforcement of content regulation 
may play a role in increasing online polarization. 
Indeed, it was recently observed that moderation policies and removal actions/bans of users produce adverse effects in terms of online polarization \cite{berger2016occasional, hughes2015isis, siegel2019online}. Users who got banned often consider this action as a badge of honor, rejoining the same social media under new identities or migrating to more tolerant platforms. The result could be either a reinforcement of their (extreme) opinion or reduced exposure to opposing voices. 
Therefore, raising awareness about the collateral costs of content policy and other interventions is crucial for making social media a less toxic environment.
\paragraph{Contribution}
This work provides a comparative analysis between two social media platforms that differ on how content moderation is applied. We select Twitter as a representative of content-regulated social media and Gab, a social network known for its willingness to ensure free speech by using little to no content moderation \cite{zannettou_2018}, like its counterpart. Despite their differences in how content policy is applied, both platforms are characterized by a similar platform design. Users are allowed to post and interact with content, together with their ability to create connections with other users. 
We perform our analysis on a timespan between $1/1/2020$ and $30/09/2020$, covering the first global wave of COVID-19. The dataset includes about three million posts and comments related to the COVID-19 topic expressed from more than one million users. We investigate consumption patterns from a user and post perspective on the two social media, assessing differences in terms of engagement. We extend this analysis by taking into account the trustworthiness of the contents published, classifying news sources accordingly to a categorization based on Media Bias/Fact Check \cite{mbfc} and NewsGuard \cite{ng}. An akin type of classification was exploited in several papers \cite{bovet2019influence, cinelli2020infodemic, cinelli2021echochambers, zollo2017debunking} bringing essential insights on the circulation of misinformation online. Therefore, we employ this dichotomy by classifying posts as \textit{Questionable} or \textit{Reliable} depending on their credibility. The same labeling was used to model the persistence of users repeatedly commenting under a post of the same outlet category.
Finally, we investigate the presence of homophily, i.e., the tendency of users to aggregate around common interests, by measuring the relationship between users and their tendency to post questionable content. We find that the content moderation imposed by Twitter promotes the existence of two echo chambers of radically different sizes. In summary, the bulk of users on Twitter seems to share and interact with verified content.

Oppositely, users on Gab show a lack of a clear preference between the two types of outlets. Questionable posts are preferred in terms of commenting persistence. However, reliable posts are more likely to be commented on as time passes. Coherently, the existence of echo chambers on Gab is not as evident as observed in the case of Twitter due to the presence of users with a relatively heterogeneous leaning.
We conclude that a valid content regulation policy produces tangible results in contrasting misinformation spreading.
\paragraph{Organization}
This work is organized as follows. Section \ref{subsec:misinfo}
describes the recent developments in the study of misinformation dynamics and their relationship with the phenomenon of polarization. Section \ref{subsec:gab} introduces the structure of Gab, providing an overview of this particular social media. Section \ref{subsec:covid_research} describes the recent advances in the study of misinformation related to COVID-19 from a social dynamics and machine learning perspective. Section \ref{sec:preliminaries} introduces the methodology and the theoretical tools applied for this study. Section \ref{sec:results} describes the results obtained from the experiments. Section \ref{sec:conclusions} provides the final remarks of this work, summarizing the results obtained and the future developments. Finally, Section \ref{sec:supporting_information} provides additional information on the results presented in the paper.
 
\section{Related Works}
\label{sec:related_works}

\subsection{Misinformation and polarization}
\label{subsec:misinfo}
The study of misinformation and the spreading of fake news has received increasing interest in recent years, disentangling the role of news consumption on mainstream and niche social media \cite{delvicario2016,Vosoughi2017falsenews,lazer2018fakenews,bovet2019influence,cinelli2020infodemic}. The presence of psychological mechanisms that affect the way users choose which news to consume \cite{Sharot2020} has been attributed to the effect of online polarization \cite{delvicario2016, vicario2019polarization} and, consecutively, to the creation of the so-called echo chambers \cite{jamieson2008echo, garrett2009echo, garimella2018political, garimella2017effect, cota2019quantifying, cinelli2021echochambers}.

\subsection{The role of Gab}
\label{subsec:gab}
Gab \cite{gab} is an online social platform that aroused much controversy in recent years. It describes itself as ``A social network that champions free speech, individual liberty and the free flow of information online. All are welcome"~\cite{gab}.
Such a claim, together with the political leaning of its founders and developers, made Gab a safe place for the alt-right movement, playing a central role in the organizations of actions to harm the offline world \cite{pittsburgh_gab}. The lack of content regulation within Gab helped the proliferation of hate speech and fake news. The risks associated with this content policy led to a series of suspensions by its former service provider and the ban of its application from online stores~\cite{zannettou_2018}.
Gab attracted the interest of researchers due to its permissive content regulatory policy and the political leaning of its users. In \textit{Lima et al.} ~\cite{lima2018inside}, authors analyzed the content shared on Gab and the leaning of users, finding a homogeneous environment prone to share right biased content. \textit{Zannettou et al.} ~\cite{zannettou_2018} characterized Gab in terms of user leanings and their contents, suggesting that this platform better suits for a safe place for right-wing extremists rather than an environment where free speech is protected. Moreover, a topological analysis performed by \textit{Cinelli et al.} ~\cite{cinelli2021echochambers} reveals that Gab users form one relevant cluster biased to the right.

Overall, all these studies suggest that Gab can be considered as a homogeneous environment where biased content and misinformation may easily proliferate.

\subsection{Recent advances in COVID-19 misinformation studies}
\label{subsec:covid_research}
Research against misinformation during the COVID-19 outbreak produced a series of results to limit the spreading of harmful information.\\ \textit{Cinelli et al.}\cite{cinelli2020infodemic} analyzed posts obtained from 5 different social media platforms (Facebook, Twitter, Instagram, Gab, and Reddit), finding out that the spreading of information is mainly driven by the peculiar structure of the social media in exam that in turn shapes the interaction patterns between users.

In the field of machine learning, \textit{Elhadad et al.}  \cite{Elhadad2020misinfocovid} introduced a framework that can identify, through a composed machine leaning approach, misleading health-related information based on a ground-truth dataset. Since their use case is related to COVID-19, the ground-truth dataset contained both epidemiological and textual data from organizations like WHO, UNICEF, UN and a range of fact-checking websites.\\
\textit{Sear et al.} \cite{sear2020opinionwar} provided a result that does not depend on a classification approach. Instead, they employed LDA-based algorithm to identify similar topics related to posts obtained from Facebook Pages belonging to pro-vax and anti-vax communities. Their findings describe how the anti-vax community develops a less focused debate on COVID-19 compared with the pro-vax counterpart. However, anti-vax seems to be more spread on the COVID-19 debate, with the result of being more positioned to attract new supporters than the pro-vax community.\\
In the context of Twitter, \textit{Jiang et al.} \cite{jiang2021social} proposed a machine-learning model based on BERT architecture which estimated user polarity within the U.S. debate by employing features related to language and network structures. They found that users belonging to the right-leaning are more active in the creation and spreading of news affiliated with their echo chamber if compared with their counterparts from the left-leaning.



\section{Preliminaries and Definitions}
\label{sec:preliminaries}
In this section, we present the methodology applied in this study. We start by introducing the data collection process of posts from Twitter and Gab together with its categorization. Then, we describe the theoretical tools behind the analysis of engagement patterns, homophily and survival lifetime.

\subsection{Data Collection}
\label{par:data_collection}

The collection of all posts concerning the COVID-19 was designed to capture its corresponding debate on social media, gathering posts and comments from both platforms. Therefore, as the first step of this process, we analyzed the most searched terms worldwide related to the aforementioned pandemic on Google Trends.  The analysis period ranges from $1/1/2020$ to $30/09/2020$. We selected four terms based on their interest and significance over time, namely: \textit{coronavirus, corona, covid, covid19}. Those terms served as a proxy, in the form of hashtags, to retrieve posts on the two social media.

The collection of Twitter posts related to the COVID-19 pandemic relied on the existence of a public dataset \cite{chen2020tracking} covering this specific topic. It includes a collection of tweet IDs, starting from $28/01/2020$, posted by accounts with a recognized influence or including representative keywords. 
To provide a categorization of the reliability of the tweets, we refined this dataset by retaining only those posts with a link, reducing the dimension to $1.1$M posts.

The same strategy was applied in the case of Gab. We queried their API to obtain posts that included at least one of the search hashtags. Due to some modifications made by the platform during the study, the API stopped providing results in chronological order since June 2020. Therefore, we started collecting all posts from the general stream until the end of the analysis period, filtering by hashtag as we previously described. This process produced a dataset of 
$\sim130$K posts containing a link.

\subsection{Questionable and Reliable Sources}
\label{par:outlet_categorization}

To evaluate the reliability of information circulating on both social media, we employed a source-based approach. We built a dataset of news outlets' domains from our dataset where each domain is labeled either as \textit{Questionable} or \textit{Reliable}. The classification relied on two fact-checking organizations called MediaBias/FactCheck (MBFC, \url{https://mediabiasfactcheck.com}) and NewsGuard (NG, \url{https://www.newsguardtech.com/}).
On MBFC, each news outlet is associated with a label that refers to its political bias, namely: \textit{Right, Right-Center, Least-Biased, Left-Center, and Left}. Similarly, the website also provides a second label that expresses its reliability, categorizing outlets as \textit{Conspiracy-Pseudoscience, Pro-Science} or \textit{Questionable}. Noticeably, the \textit{Questionable} set includes a wide range of political biases, from \textit{Extreme Left }to \textit{Extreme Right}. For instance, the \textit{Right} label is associated with Fox News, the \textit{Questionable} label to Breitbart (a famous right extremist outlet), and the \textit{Pro-Science} label to \textit{Science}.
MBFC also provides a classification based on a \textit{ranking bias score} that depends on four categories: \textit{Biased Wording/Headlines, Factual/Sourcing, Story Choices,} and \textit{Political Affiliation}. Each category is rated on a $0-10$ scale, with $0$ indicating the absence of bias and $10$ indicating the presence of maximum bias. The \textit{bias outlet score} is computed as the average of the four score categories. Likewise, NG classifies news outlets into four categories based on nine journalistic criteria, each of them having a specific score whose sum ranges between $0$ and $100$. Outlets with a score of at least $60$ points are considered compliant with the basic standards of credibility and transparency. Otherwise, they are recognized as outlets that lack of credibility. A different characterization is provided for humor and platforms websites, not accounting for the categorization process.

Given the different ways of classifying information sources from the two organizations, the following heuristic was applied. On MBFC, all the outlets already classified as \textit{Questionable} or belonging to the category \textit{Conspiracy-Pseudoscience} were labeled as \textit{Questionable}. The remaining categories were labeled as \textit{Reliable}. Coherently, outlets on NG were classified based on their score, maintaining the dichotomy 
provided by the website. We choose a score of 60 as threshold to consider an outlet as \textit{Reliable} (score $>60$), otherwise it is referred as \textit{Questionable} (score $\le60$).

Considering a total of $2738$ news outlets provided by the two organizations, $2701$ belonging to MBFC and $37$ to NG, we end up with $814$ outlets classified as Questionable and $1924$ outlets classified as Reliable.

\subsection{User Leaning}
\label{par:user_leaning}
To measure the extent to which a user is associated with the consumption of questionable or reliable contents, we introduce the \emph{user leaning} $q$. We define it in the range $q \in [0,1]$,  where $0$ means that a user posts contents exclusively associated with reliable sources, and $1$ means that a user puts into circulation only questionable posts. 

Formally, the user leaning can be defined as follows: let $\mathcal{P}$ be the set of all posts with a URL matching a domain in our dataset and $\mathcal{U}$ the set containing all the users with at least a categorized post. At each element $p_j \in \mathcal{P}$ is associated a binary value $l_j \in \{0,1\}$ based on the domain of the link contained: if the URL refers to a domain classified as questionable then $l_j=1$, otherwise $l_j=0$. Considering a user $u_i$ in a bipartite network between users and posts, then the user leaning $q_i$ of a user $u_i$ can be defined as:
\begin{equation}
    q_i=\frac{1}{k_i}\sum_{j=1}^{k_i} l_j \enspace ,
    \label{eq:user_leaning}
\end{equation}
where $l_j$ is the leaning score of the j-th neighbor of the user $u_i$, and $k_i$ is the number of categorized contents that the user posted.

\subsection{Comparison of power law distributions}
\label{par:statistical_tests}
Most quantities related to the activity of users on social media show a heavy tailed distribution of discrete variables. Given the discrete nature of such distributions, we could not rely on Kolmogorov-Smirnov \cite{clauset2009powerlaw} test to assess whether two distributions present significant differences between each other. Indeed, such a test assumes that distributions must be continuous, and the presence of a large number of ties in the long-tailed distributions that we want to compare may lead to the computation of biased p-values. To overcome this issue, we employed a methodology proposed in \textit{Zollo et al.}\cite{zollo2017debunking} which makes use of a Wald Test \cite{wald1943test} to assess significant differences between the scaling parameters of two long-tailed distributions.   

\begin{figure*}
    \centering
    \includegraphics[trim = 0mm 0mm 0mm 0mm, scale = 0.2]{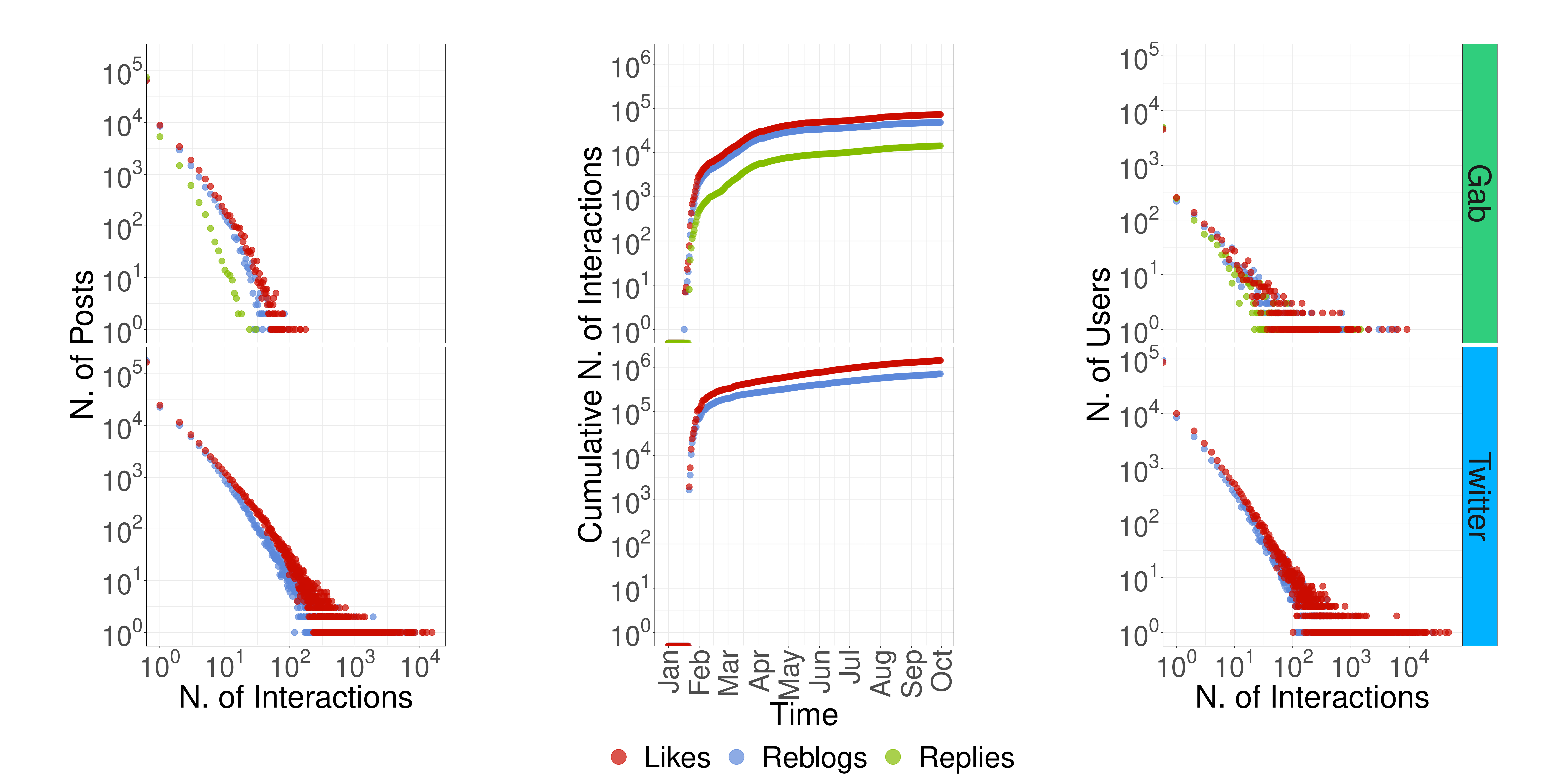}
    \caption{Representation of the engagement collected on Gab (upper panel) and Twitter (bottom panel).
    \textit{Left column}: frequency distribution of the interactions for posts, defined as \textit{Likes}, \textit{Reblogs} (or \textit{Retweets)} and \textit{Replies}. A like is generally considered positive feedback on a news item. A reblog indicates a desire to spread a news item to friends. A reply can have multiple features and meanings and can generate collective debate. Both social media shows a heavy-tailed distribution that allows room for large deviations, i.e., some posts go viral. 
    \textit{Middle column}: evolution of the cumulative number of interactions over time. The general trend shows a rapid increase during February 2020, in parallel with the spreading of the COVID-19 outbreak. The absence of replies on Twitter is due to the limitations provided by their API. 
\textit{Right column:} frequency distribution of interactions received by users. Similarly to posts, the distribution is heavy-tailed, describing how users tend to collect similar values of different interactions as their number increases. }
\label{fig:overall_attention_pattern_general}
\end{figure*}

\subsection{Kaplan-Meier Estimator for lifetime analysis}
\label{par:km_definition}
Let $T \in [0, +\infty]$ be a random variable which represents the time an event takes place. The probability that a randomly selected subject lives up to time $t$ is called Survival Function $S(t) = P(T \le t)$. The estimation of this probability is provided by the Kaplan-Meier estimator \cite{kaplanmeier1958}, defined as
\begin{equation}
    \hat{S}(t) = \prod_{t_i \le t}{\left(1 - \frac{ d_{i}}{n_{i}}\right)},
    \label{eq:kaplan_meier_estimator}
\end{equation} where $d_i$ is the number of events that happened at time $t_i$ and $n_i$ are the numbers of subjects who survived up to time $t_i$. 
In user lifetime analysis, we define as $d_i$ the number of users who have been commenting on a post up to $t_i$ days and $n_i$ the number of users who stopped commenting after $t_i$ days. Similarly, in post lifetime analysis, we define as $d_i$ the number of posts that have been receiving a comment up to $t_i$ days and $n_i$ the number of posts that stopped receiving comments at $t_i$ days. In both cases, our $N$ distinct events times $t_1, \small t_2,\small \dots,\small t_N \ge 0$ are independent, which is required by the assumptions of the Kaplan-Meier estimator.

\paragraph{Homophily Analysis in User Following Networks}
\label{par:homophily}
Results from the computation of the user leaning $q$ were generalized to measure user homophily based on his news consumption. This phenomenon was modeled from a network perspective based on the work originally proposed by \textit{Cota et al.} \cite{cota2019echo}. We considered a new network in which a user is represented as a node $i$ with a given leaning $q_i$. Each node can be connected to others through a \textit{following} relationship: if user $i$ follows user $j$ on the social media in exam, their corresponding representation in the adjacency matrix $A$ is $A_{ij} = 1$, meaning that there is a directed edge between $i$ and $j$. In case of no relationship between the two users, we have $A_{ij} = 0$. This representation allows the computation of a measure called \textit{average neighborhood leaning} that quantifies the mean leaning from all those users followed by a given one. It is defined as
\begin{equation}
    q_i^N = \frac{1}{k_i^{\rightarrow}}\sum_jA_{ij}q_{j}\enspace,
    \label{eq:average_neighborhood_leaning}
\end{equation}

where $k_i^{\rightarrow} = \sum_jA_{ij}$ is the out-degree of node $i$, i.e., the number of users followed by user $i$.

\begin{figure*}
    \centering
    \includegraphics[trim = 0mm 0mm 0mm 0mm, scale = 0.2]{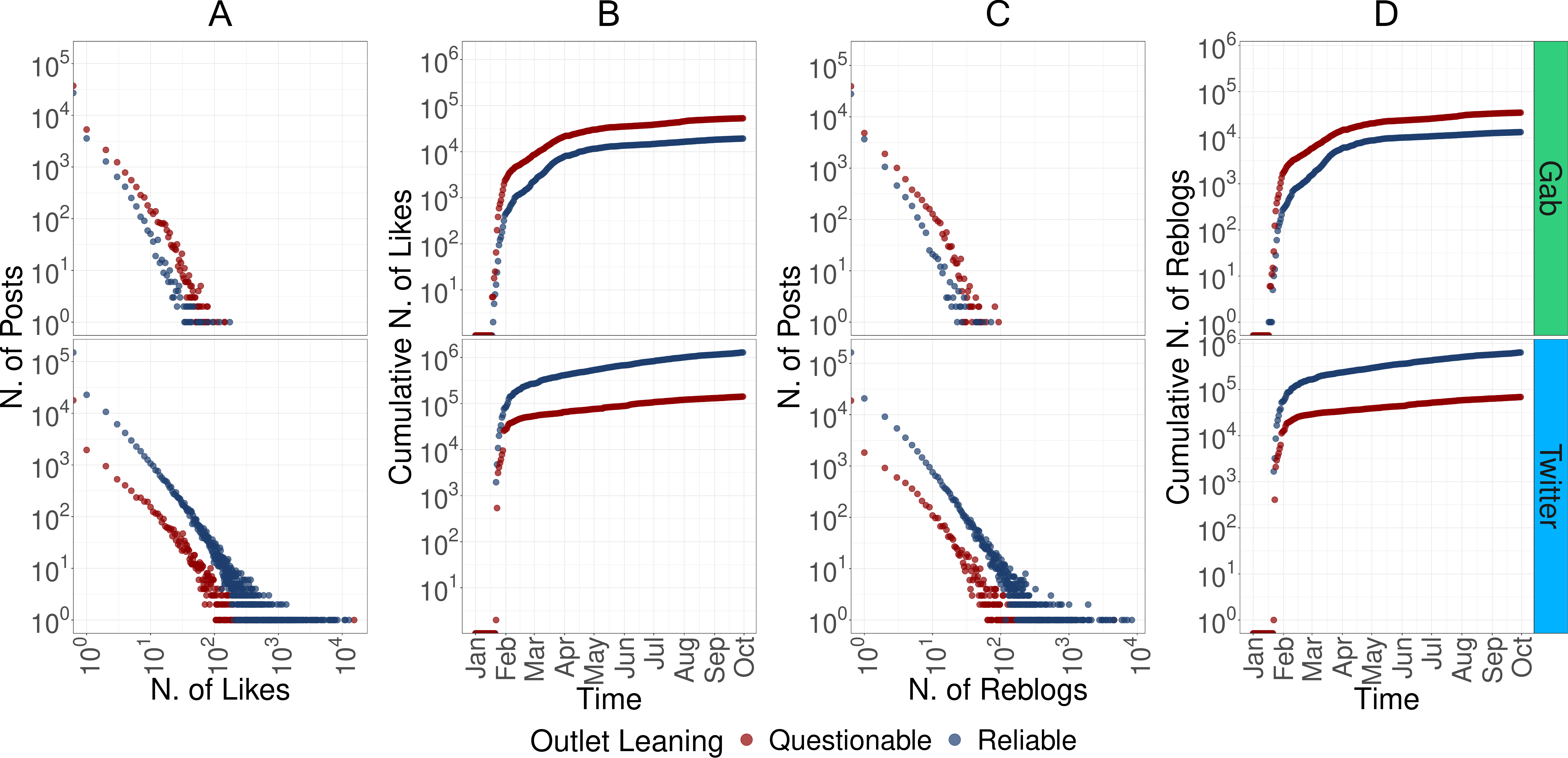}
    \caption{
    \textit{Column A-B}: categorized distribution of the number of posts
    against the number of likes they received with its cumulative evolution. The distributions show some evidence about content preference on both platforms. Users on Gab show signs of better appreciation toward questionable posts, supported by the lack of clear content regulation. Twitter, oppositely, shows strong evidence towards the appreciation of reliable content, with a remarkable gap between the two categories. From a cumulative perspective, the regulation imposed by Twitter results in an increasing divergence between questionable and reliable posts, showing how the latter category is the most preferred one. The same does not apply to Gab, whose divergence seems not to increase during the analysis period. 
    \textit{Column C-D:} categorized distribution of the number of posts against the number of reblogs or retweets they received with its cumulative evolution. The previous considerations also apply to this kind of interaction, describing the willingness of users to inject the contents they support into the news feed of their followers.}
    \label{fig:post_attention_pattern_categorized}
\end{figure*}

\section{Result and Discussion}
\label{sec:results}

This work aims at performing a comparative analysis of two social media, namely Twitter and Gab, in order to understand how news consumption and social dynamics change in presence of two radically different types of content regulation policies (more stringent in the case of Twitter, almost absent in the case of Gab). The following results will provide insights to explain this behavior from different perspectives. At first, we analyze the engagement of users with posts, which we consider as separated into two categories named questionable and reliable. Then, we quantify the commenting behavior of users and posts. Lastly, we provide a network analysis to measure the tendency of users to aggregate with like-minded peers, describing how the presence of content regulation may be correlated with the polarization towards specific narratives.

\subsection{Consumption Patterns}
\label{subsec:consumption_patterns}
We investigate how the engagement on the two social media differs in relationship with the COVID-19 topic. Fig.~\ref{fig:overall_attention_pattern_general} compares the engagement distribution for posts and users. Despite the differences in terms of scale that are attributable to the size of the platforms' user base, we observe that both frequency distributions are long-tailed. This feature provides a first evidence in the consumption of news, showing that interaction patterns are similar regardless of the content moderation imposed.

\begin{figure}
    \centering
    \includegraphics[trim = 65mm 0mm 0mm 0mm, scale = 0.18]{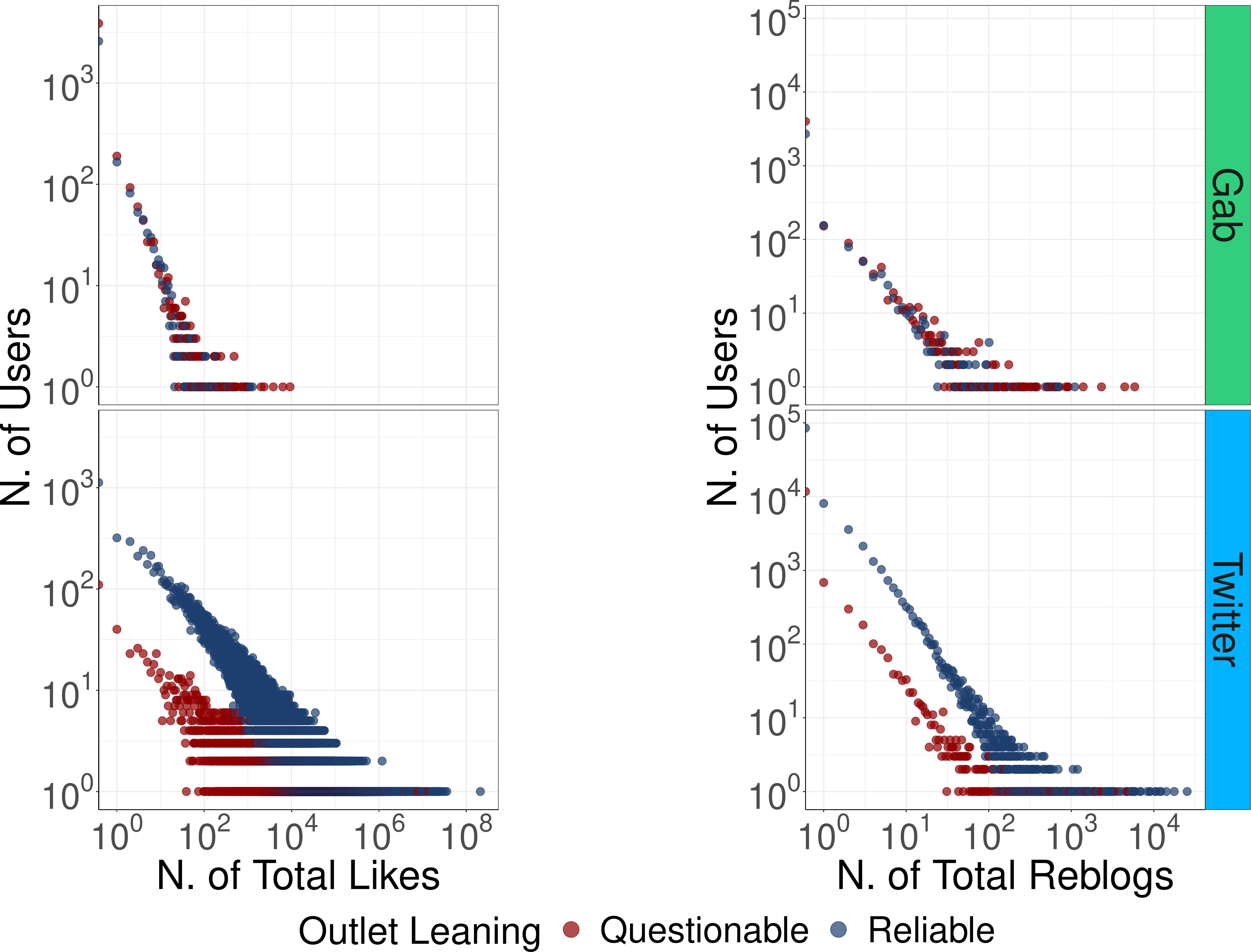}
    \caption{Distribution of likes (left column) and reblogs (right column) received by users posting Questionable or Reliable contents on Gab (upper panel) and Twitter (bottom panel). The figure shows how the presence of content regulations, performed by Twitter, results in a greater appreciation towards users who post reliable content. Gab, instead, shows a mixed endorsement pattern in which the appreciation towards users does not depend on the category of the content they post.}
    \label{fig:user_attention_pattern_categorized}
\end{figure}

Next, we extend the analysis of consumption patterns by categorizing posts, based on their outlet leaning, into Questionable or Reliable. The resulting distributions from the application of this dichotomy are represented in Fig. \ref{fig:post_attention_pattern_categorized} and \ref{fig:user_attention_pattern_categorized}. Fig. \ref{fig:post_attention_pattern_categorized} displays the distribution of the number of likes and shares (reblogs or retweets) obtained by posts in our dataset, together with the corresponding cumulative. Similarly, Fig. \ref{fig:user_attention_pattern_categorized} describes the frequency distribution of the same kind of interactions from a user perspective. In general, we observe how Twitter users show a larger appreciation of reliable posts, establishing a clear gap from questionable ones that increases during the analysis period. This difference can be attributed to the commitment of Twitter to limit the spreading of unverified contents \cite{twitter_misinfo}. The opposite scenario happens on Gab, in which the consumption patterns do not show a clear sign of polarization towards a specific kind of narrative. This provides some evidence of how users belonging to segregated environments like Gab are not interested in the origin of the content itself. Instead, they tend to self-segregate within environments in which they can consume and spread questionable content. Therefore, the lack of regulation on this platform may allow them to perform information operations \cite{weedon2017information}, i.e., a category of actions
taken by organized actors (governments or non-state actors) to distort domestic or foreign
political sentiment, against other users who do not share the mainstream system of beliefs of the community.

In order to assess the similarity between the distributions deriving from the consumption patterns of questionable and reliable posts, we fit power-law distributions to such data and perform a statistical evaluation of their scaling parameters using the Wald test. For Gab, all the obtained p-values were significantly higher than $0.05$, describing how questionable and reliable distributions are comparable in terms of the engagement produced. The same behavior is found on Twitter, except for the likes distribution whose p-value is less than $0.001$, describing a significant difference in the way users
engage with questionable and reliable content. 


We can conclude that the presence of content moderation is associated with a significant reduction of the flowing of misinformation. The avoidance of these countermeasures, as reported on Gab, seems to be associated with more heterogeneous news consumption in terms of outlet category. This particular shape of the news diet may be exploited to conduct information operations. Statistical tests indicate how questionable and reliable contents produce similar engagement behaviors within the same social media. In the end, such moderation helps the emergence of segregation, a condition in which users are affected by the restriction applied in terms of accessibility to the contents they are affiliated with but not in their engagement behaviors.

\subsection{Characterizing Commenting Behavior for Questionable and Reliable posts}
\label{subsec:commenting_lifetime}
\begin{figure}
    \centering
    \includegraphics[trim = 20mm 0mm 0mm 0mm, scale = 0.2]{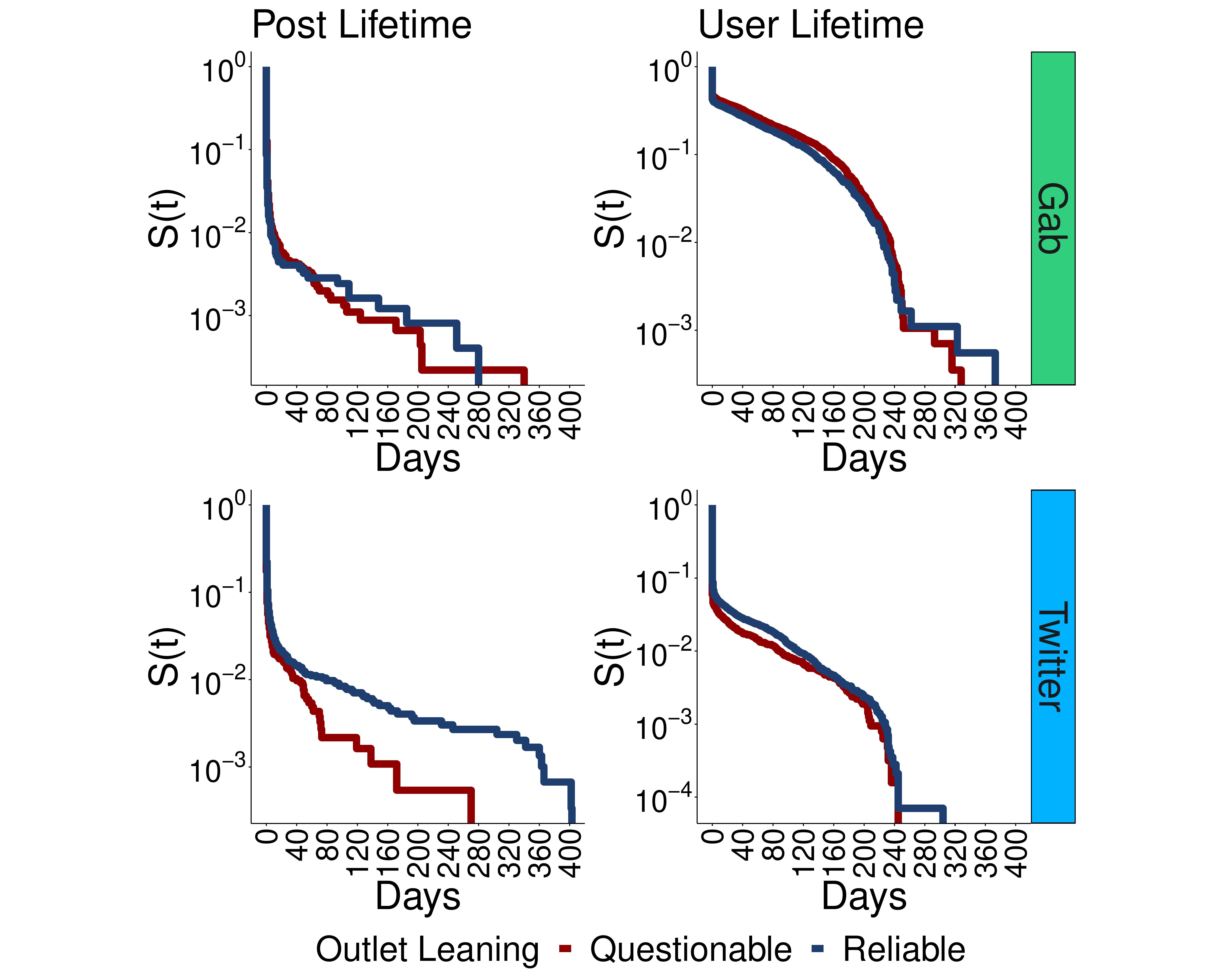}
    \caption{Kaplan-Meier estimates for Gab (upper panel) and Twitter (lower panel), grouped by outlet category.\\
\textit{Left column}: estimates obtained through the computation of post lifetime, i.e., the period between the first and last comment a post received. \textit{Right column}: estimates obtained through the computation of post lifetime, i.e., the period between the user's first and last comment.\\
Gab shows how the lack of content regulation is associated with a commenting behavior that underlines a preference towards questionable content. This behavior is characterized by a discrepancy between the outlet category with the highest commenting persistence both on user and post lifetimes. By contrast, the introduction of content policies from Twitter makes reliable content those with the highest commenting persistence, which does not depend on the lifetime perspective.}
    \label{fig:km_lifetime}
\end{figure}
To quantify the persistence of comments concerning users and posts, we employed Kaplan-Meier estimates of two survival functions. The first accounts for the period between the first and last comment received from posts. The second instead considers the period between the first and last comment made by a user. To characterize any significant difference in the two survival functions, we perform the Peto \& Peto \cite{peto1972test} test. The upper panel of Figure~\ref{fig:km_lifetime} shows the Kaplan-Estimates computed on Gab, grouped by outlet category. The test performed on its post and user lifetimes produces a p-value of $0.026$ and $0.001$, respectively.  Therefore, we can conclude that the commenting persistence on Gab may be subjected to the outlet category of the post commented. Indeed, post lifetime on questionable posts reports a lower probability of being commented as time increases despite its longer persistence, reaching a maximum $340$ days. Results from user lifetime estimation, instead, describe how users are more likely to comment on questionable posts for the first $240$ days after post creation. After that time, the survival probability becomes higher on reliable posts.\\
In the end, we can conclude that the commenting behaviors on Gab reflect the general leaning of its community. Users are more likely to comment on questionable posts since their contents adhere to a common system of beliefs oriented to conspiracy theories. Coherently, the significant commenting persistence reported on reliable posts may describe the desire of users to express their dissent against the narratives introduced from such posts.

Next, we examine the commenting persistence on Twitter. Results from Peto \& Peto test on the post and user lifetimes report a p-value equal to $0.011$ and $0.0055$ respectively, stating how the survival functions on both lifetimes differentiate with respect to the outlet category of the posts commented. Indeed, such estimations on Twitter describe a uniformity in the commenting behavior for the reliable category. This fact also provides further evidence about how content moderation can discourage users from expressing their views under posts whose authority is not verified. 

In summary, Gab demonstrates how the lack of content policy helps the emergence of the narratives that characterize this environment, resulting in a discrepancy between the outlet categories with the most commenting persistence on the two lifetimes. However, when the content policy is applied, like on Twitter, such discrepancy dissolves, resulting in a commenting behavior that favors reliable content.

\subsection{Quantifying Polarization}
\label{subsec:polarization}
\begin{figure}%
    \centering
    \subfloat[\centering Twitter]{{\includegraphics[trim = 0mm 0mm 0mm 0mm, scale = 0.3]{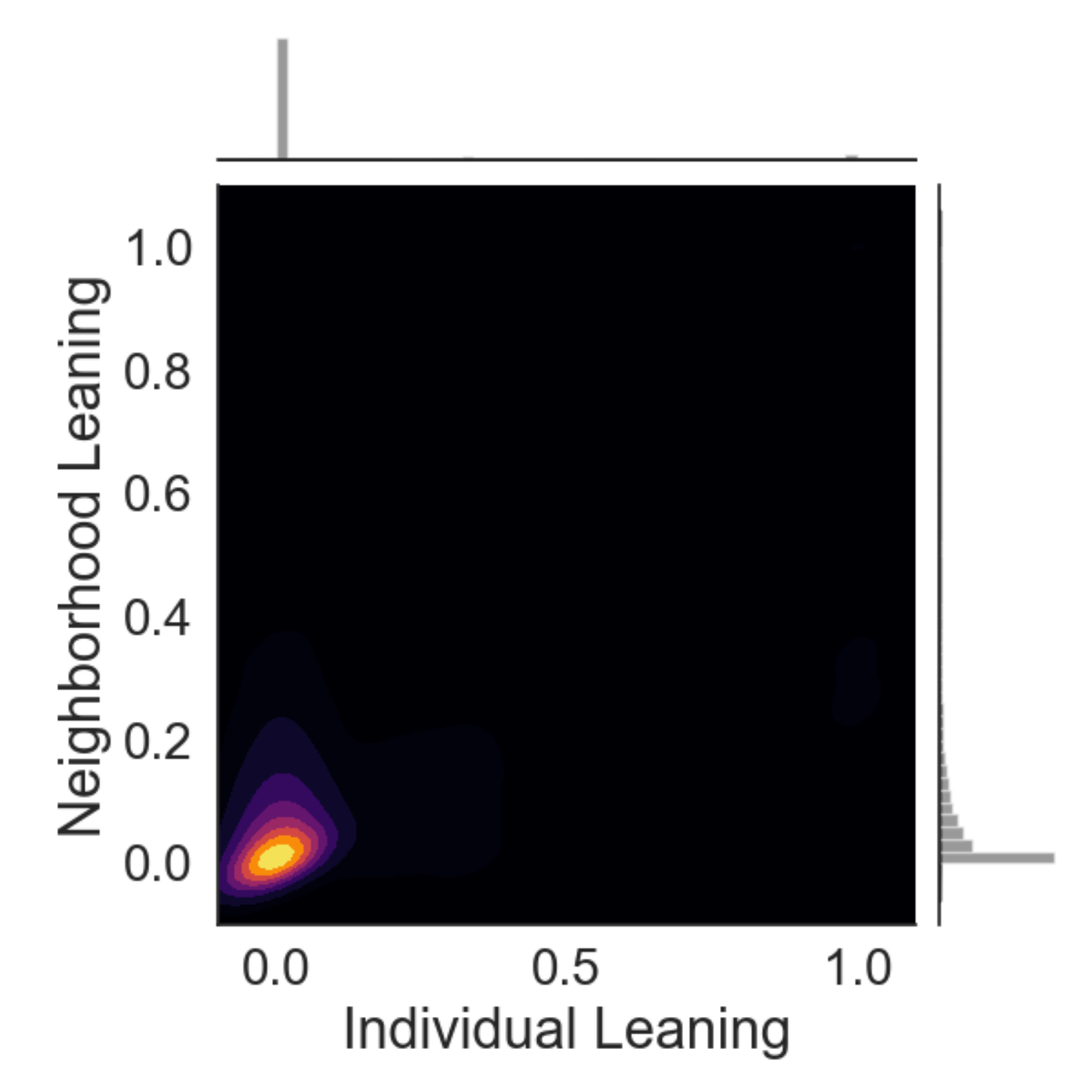} }
    \label{fig:echo_chambers_twitter}
    }%
    \subfloat[\centering Gab]{{\includegraphics[trim = 10mm 0mm 0mm 0mm, scale = 0.3]{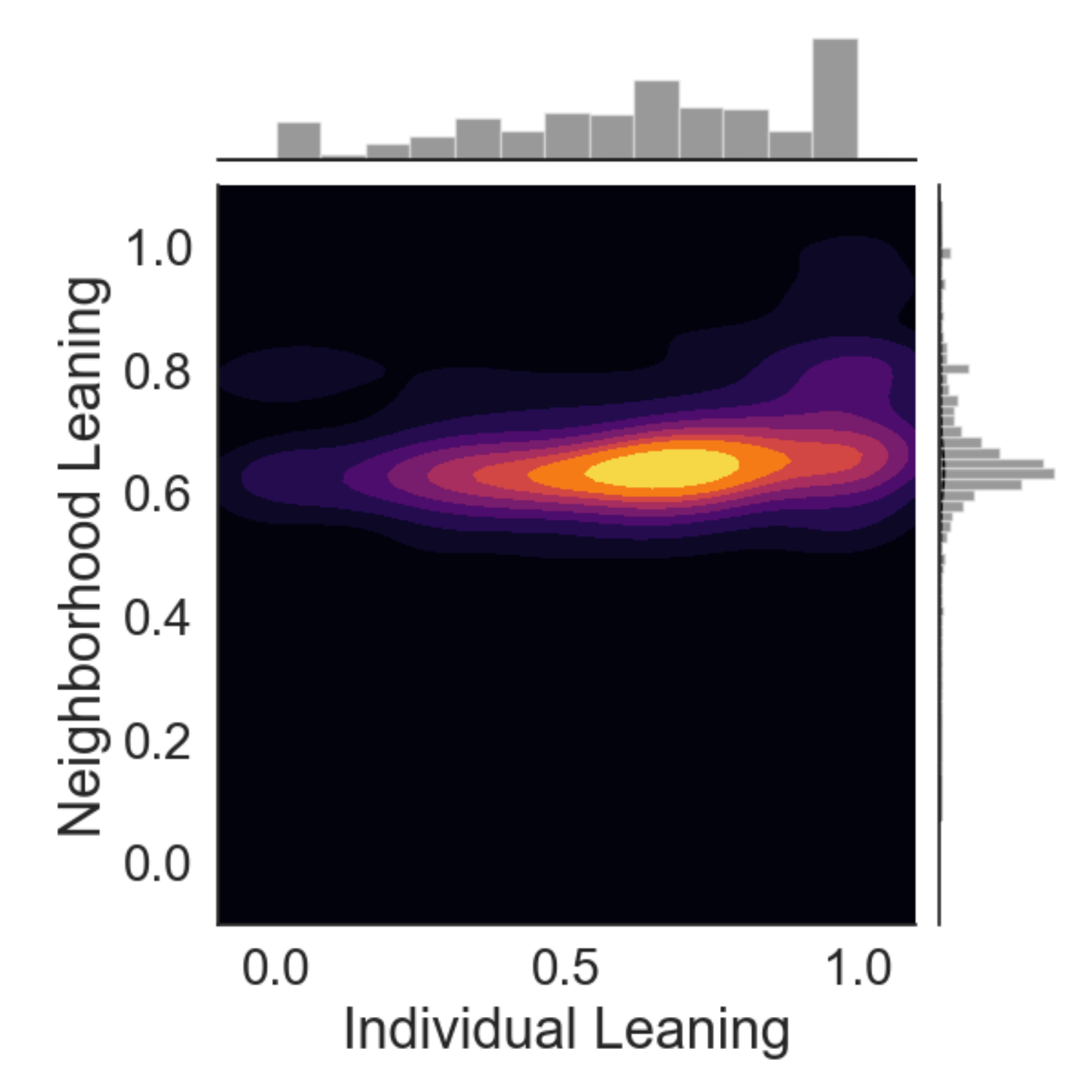} }
        \label{fig:echo_chambers_gab}
        }%
    \caption{Joint distribution between individual and average neighborhood leaning of all users posting classifiable contents at least three times on Twitter (left) and Gab (right). The figure shows further evidence about the regulation imposed by Twitter which results in the creation of a unique echo chamber of users with strong posting habits towards reliable content. Oppositely, Gab shows the presence of an echo chamber in which both individual and neighborhood leanings are concentrated around high values of the intervals, with a greater dispersion due to the mixed posting habits of users.
    }%
    \label{fig:echo_chambers}%
\end{figure}

The presence of content moderation may affect how users develop homophily, i.e., the tendency to surround themselves with other peers who share the same narratives or system of beliefs. To quantify this phenomenon, we build a network in which the nodes represent the users  $i$ with their corresponding leaning $x_i$, while the edges represent the \textit{following} relationship with other users that occurs on the social media. This representation allows us to measure the neighborhood leaning $x_i^{N}$, i.e., a measure of the characteristic leaning of the network surrounding user $i$. Figure \ref{fig:echo_chambers} displays the joint distribution between the individual leaning of a user $x_i$ and its corresponding neighborhood leaning $x_i^N$, on Twitter and Gab. In addition to this, the marginal probability distributions $P(x)$ and $P^{N}(x)$, referring to the individual and average neighborhood leaning, are represented on their corresponding axis. Lastly, the density of users at point $(x, x^N)$ is represented as a contour map: the brighter the color in that point, the higher the user density. Results described in Figure \ref{fig:echo_chambers_twitter} show the presence of homophily on Twitter, characterized by a strong correlation of leanings in correspondence of low values. The existence of a second echo chamber of incomparable size made of users with high individual leaning, and therefore not represented in the main figure but only visible in the marginal distributions, signals strong segregation between two communities. This finding also indicates how content regulations actively affect the shape of the news diet of users in the context of the COVID-19 pandemic. Indeed, the concentration around small values for both leanings provides evidence about the effectiveness of the moderation imposed by the platform against posts and users that promote questionable content. On the other side, Gab shows a more heterogeneous behavior, as represented in Figure \ref{fig:echo_chambers_gab}. Indeed, the joint distribution spreads over different values of the individual leaning domain, with the highest mode represented in correspondence of the point $(0.6, 0.6)$. We observe that on average users, regardless of their leaning, are surrounded by a neighborhood skewed towards questionable contents. Only very few users have a reliable-based leaning, who are also likely to be those with a weaker activity since they could be on Gab just for curiosity or dissing. Furthermore, the outlet category of the news that users post is not relevant anymore since the user's peers share a leaning with a high value. Finally, these findings may suggest that questionable news is employed to support the narrative of the environment, whilst reliable ones are only used to perform information operations by changing the original meaning of the posts through a comment. 

\section{Conclusions}
\label{sec:conclusions}
In this work, we compared two social media, Twitter and Gab, to investigate the interplay between content regulation policies and news consumption. We provide quantitative measures of such differences by evaluating the engagement of users and posts. These measures are then extended by providing a categorization of news outlets. Next, we measure the commenting persistence of users and posts to describe their ability to express themselves under posts belonging to a specific outlet category. In the end, we characterize the presence of homophily, investigating how users with a specific leaning are more likely to surround themselves with users who share the same narratives.

Our results show how the application of content regulation, performed by Twitter, limits the diffusion of fake news and conspiracy theories, shaping the news consumption and the polarization of users towards reliable content. The avoidance of these countermeasures, carried on by Gab, provides results that underline the presence of patterns related to information operations. Indeed, users tend to engage with questionable and reliable content comparably. However, their commenting behavior and the assessment of the homophily in this environment describe a systematic affiliation towards questionable contents.

We conclude that content policies cover an important role against the circulation of harmful content, especially in the context of the COVID-19 pandemic. Our work provides meaningful evidence in this direction, indicating how a lack of content policy is associated with the emergence of harmful narratives that promote questionable content and mistrust everything that goes against them.

Future implementations of this work may focus on the dissing/endorsement behavior promoted by users in segregated environments like Gab, analyzing those mechanisms from a textual perspective. Furthermore, a topological analysis of users who perform information operations in such environments may be relevant to understand their inner dynamics and to promote specific countermeasures.

\appendices

\ifCLASSOPTIONcaptionsoff
  \newpage
\fi



\bibliographystyle{IEEEtran}
\bibliography{bare_tran}

%

\clearpage

\section{Supporting Information}
\label{sec:supporting_information}
Here we provide further details about the datasets employed for the study, the cumulative and daily evolution of posts and new users posting and the numerical results of the statistical tests performed. The description of the dataset is reported in Section \ref{subsec:data_breakdown}, the representation of the time series is reported in Section \ref{subsec:time_series} while the results of the statistical test performed are described in Section \ref{subsec:test_results}.

\subsection{Data breakdown}
Here we report the composition of the dataset for Gab and Twitter employed for the study, described in Table \ref{tab:gab_data} and Table \ref{tab:twitter_data} respectively. Each dataset represents the number of unique posts, users and comments, as well as their engagement quantities over the different data collection and processing steps. Due to Twitter API limitations, posts were initially gathered without any information about their number of comments. The collection of this quantity was performed after the categorization of the news outlets through the employment of specific APIs provided from Twitter as part of its Academic Research Program \cite{twitter_academic}.

\label{subsec:data_breakdown}
\begin{table*}[]
\centering
\begin{tabular}{|l|l|l|l|l|l|}
\hline
\textbf{Category}          & \textbf{Overall}                & \textbf{Containing search hashtags }     & \textbf{Categorized}                    & \textbf{Questionable}                                       & \textbf{Reliable}                                  \\ \hline
\textbf{Number of Posts }           & 205 458 & 130 864 & 83 784 & 49 772                     & 34 012                     \\ \hline
\textbf{Number of users}   & 11 063  & 8 194   & 5 681  & 4 660                      & 3 289                      \\ \hline
\textbf{Number of Likes}   & 234 255 & 117 281 & 72 435 & 53 154                     & 19 281                     \\ \hline
\textbf{Number of Reblogs} & 138 793 & 75 250  & 48 172 & 34 960                     & 13 212                     \\ \hline
\textbf{Number of Comments} & 42 993  & 22 287  & 14 165 & \multicolumn{1}{l|}{9 489} & \multicolumn{1}{l|}{4 676} \\ \hline 
\end{tabular}
\caption{Composition of Gab Dataset where each column represents the quantities of collected posts collected during the preprocessing phase. }
\label{tab:gab_data}
\end{table*}

\begin{table*}[]
\centering
\begin{tabular}{|l|l|l|l|l|l|}
\hline
\textbf{Category}          & \textbf{Overall} & \textbf{Containing search hashtags} & \textbf{Categorized} & \textbf{Questionable} & \textbf{Reliable}  \\ \hline
\textbf{Number of Posts}            & 2 668 286        & 1 110 030                  & 244 430     & 25 121       & 219 309   \\ \hline
\textbf{Number of users}   & 1 185 541        & 382 449                    & 118 635     & 13 711       & 108 153   \\ \hline
\textbf{Number of Likes}   & 18 610 555       & 5 885 562                  & 1 422 629   & 141 273      & 1 281 356 \\ \hline
\textbf{Number of Reblogs} & 7 753 971        & 2 862 098                  & 703 765     & 68 609       & 635 156   \\ \hline
\textbf{Number of Comments} & NA               & NA                         & 30 262              &  8771    & 21 491    \\ \hline
\end{tabular}
\caption{Composition of Twitter Dataset where each  column represents the quantities of collected posts collected during the preprocessing phase. }
\label{tab:twitter_data}
\end{table*}

\subsection{Time Series Evolution}
\begin{figure}
    \centering
    \includegraphics[scale = 0.18]{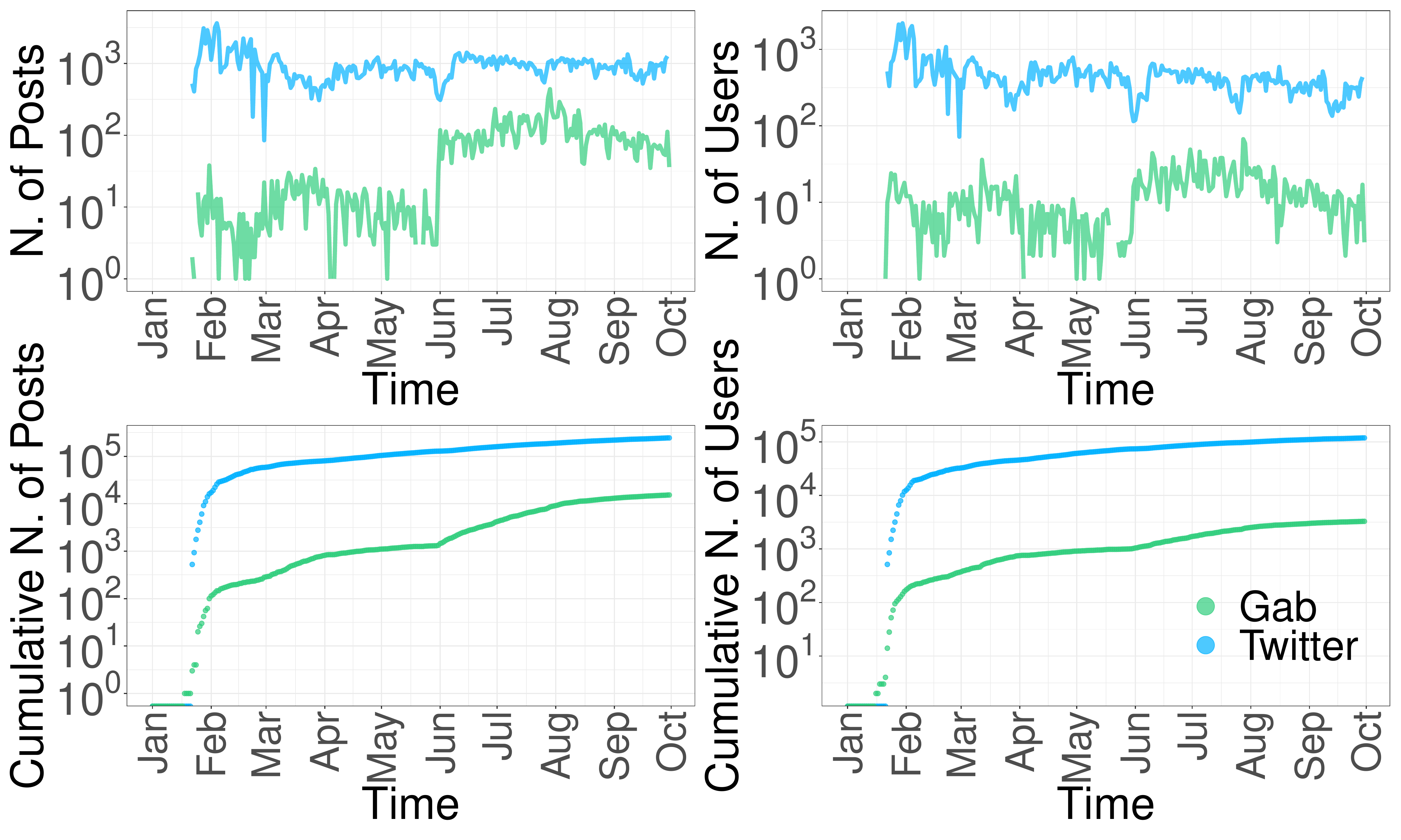}
    \caption{\textit{Upper panel:} Time series evolution of new posts (left) and users posting for the first time (right) on Gab and Twitter. \textit{Lower panel:} cumulative evolution of new posts (left) and users posting of the first time (right) on Gab and Twitter. }
    \label{fig:time_series_overall}
\end{figure}

\begin{figure}
    \centering
    \includegraphics[scale = 0.18]{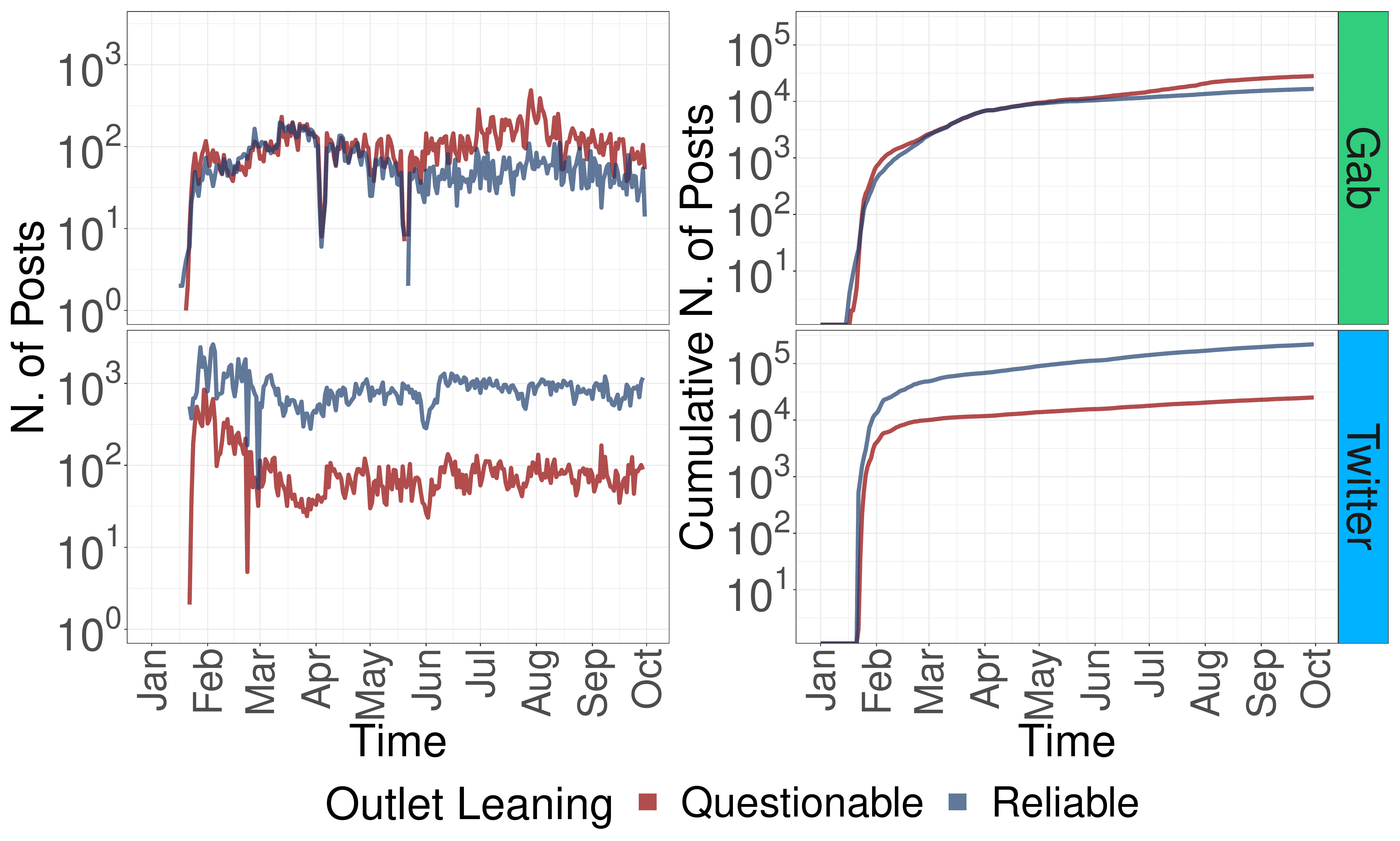}
    \caption{Evolution of posts based on their outlet leaning, categorized as \textit{Questionable} or \textit{Reliable}.\\
    \textit{Upper panel:} Time series evolution of new posts (left) together with its cumulative representation (right) on Gab.\\
    \textit{Lower panel:} Time series evolution of new posts (left) together with its cumulative representation (right) on Twitter.}
    \label{fig:post_time_series_categorized}
\end{figure}

Here we report the evolution of posts and new users collected on Twitter and Gab during the analysis time. Figure \ref{fig:time_series_overall} shows the overall evolution of such quantities, while Figure \ref{fig:post_time_series_categorized} shows the previous evolution after performed the categorization of the news outlets.
Figure \ref{fig:time_series_overall} describes a spike on Gab in the number of posts and users in correspondence of June. This was due to the change of the collecting method. Indeed, the lack of chronological order of posts reported on Gab APIs since June required the gathering of all posts from the general stream, which was then filtered by the search hashtags in order to be compliant with the data collection process.
\label{subsec:time_series}

\subsection{Results of Comparison between power law distributions}
\label{subsec:test_results}
Here we report the information related to the comparison of the power law fits described in Section \ref{subsec:consumption_patterns} by means of the Wald test. Tables \ref{tab:gab_post_attention_patterns_distrib} - \ref{tab:twitter_user_attention_patterns_distrib} report the estimated coefficients of each power law fit, i.e., $\hat{\alpha}$ and $\hat{x}_{\mathrm{min}}$, depending on the engagement and news outlet category. Furthermore, the results of the Wald test score applied on the previous engagement categories are reported, together with the corresponding p-values. 

\begin{table*}[]
\centering
\begin{tabular}{|l|l|l|l|l|l|}
\hline

\multicolumn{3}{|c|}{\textbf{Likes}}                    & \multicolumn{3}{c|}{\textbf{Reblogs}}                     \\ \hline
                & $\hat{\alpha}$ & $\hat{x}_{min}$ &                   & $\hat{\alpha}$ & $\hat{x}_{min}$ \\ \hline
Questionable    & 1.32         & 1             & Questionable      & 1.31         & 1             \\ \hline
Reliable        & 1.37         & 1             & Reliable          & 1.33         & 1             \\ \hline
Wald test score &              & 0.22          & Wald's test score &              & 0.034         \\ \hline
p-value         &              & 0.64          & p-value           &              & 0.85          \\ \hline
\end{tabular}
\caption{Power law fits of Likes and Reblogs for post consumption patterns on Gab, together with the result of Wald test between the scaling parameters of each outlet category.}
\label{tab:gab_post_attention_patterns_distrib}
\end{table*}

\begin{table*}[]
\centering
\begin{tabular}{|l|l|l|l|l|l|}
\hline
\multicolumn{3}{|c|}{\textbf{Likes}}                        & \multicolumn{3}{c|}{\textbf{Reblogs}}                         \\ \hline
                & $\hat{\alpha}$ & $\hat{x}_{min}$ &                   & $\hat{\alpha}$ & $\hat{x}_{min}$ \\ \hline
Questionable    & 1.52           & 1               & Questionable      & 1.57           & 1               \\ \hline
Reliable        & 1.55           & 1               & Reliable          & 1.55           & 1               \\ \hline
Wald test score &                & 0.15            & Wald's test score &                & 0.05            \\ \hline
p-value         &                & 0.70            & p-value           &                & 0.83            \\ \hline
\end{tabular}
\caption{Power law fits of Likes and Reblogs for post consumption patterns on Twitter, together with the result of Wald test between the scaling parameters of each outlet category.}
\label{tab:twitter_post_attention_patterns_distrib}
\end{table*}

\begin{table*}[]
\centering
\begin{tabular}{|l|l|l|l|l|l|}
\hline
\multicolumn{3}{|c|}{\textbf{Likes}}                        & \multicolumn{3}{c|}{\textbf{Reblogs}}                         \\ \hline
                & $\hat{\alpha}$ & $\hat{x}_{min}$ &                   & $\hat{\alpha}$ & $\hat{x}_{min}$ \\ \hline
Questionable    & 1.81           & 1               & Questionable      & 1.83           & 1               \\ \hline
Reliable        & 1.83           & 1               & Reliable          & 1.83           & 1               \\ \hline
Wald test score &                & 0.015           & Wald's test score &                & 0.0004          \\ \hline
p-value         &                & 0.90            & p-value           &                & 0.98            \\ \hline
\end{tabular}
\caption{Power law fits of Likes and Reblogs for user consumption patterns on Gab, together with the result of Wald test between the scaling parameters of each outlet category.}
\label{tab:gab_user_attention_patterns_distrib}
\end{table*}

\begin{table*}[]
\centering
\begin{tabular}{|l|l|l|l|l|l|}
\hline
\multicolumn{3}{|c|}{\textbf{Likes}}                        & \multicolumn{3}{c|}{\textbf{Reblogs}}                         \\ \hline
                & $\hat{\alpha}$ & $\hat{x}_{min}$ &                   & $\hat{\alpha}$ & $\hat{x}_{min}$ \\ \hline
Questionable    & 3.33           & 1               & Questionable      & 1.75           & 1               \\ \hline
Reliable        & 2.21           & 1               & Reliable          & 1.67           & 1               \\ \hline
Wald test score &                & 1286.34         & Wald's test score &                & 1.06            \\ \hline
p-value         &                & $<$ 0.001         & p-value           &                & 0.30            \\ \hline
\end{tabular}
\caption{Power law fits of Likes and Reblogs for user consumption patterns on Twitter, together with the result of Wald test between the scaling parameters of each outlet category.}
\label{tab:twitter_user_attention_patterns_distrib}
\end{table*}

\clearpage
\begin{IEEEbiography}[{\includegraphics[width=1in,height=1.25in,clip,keepaspectratio]{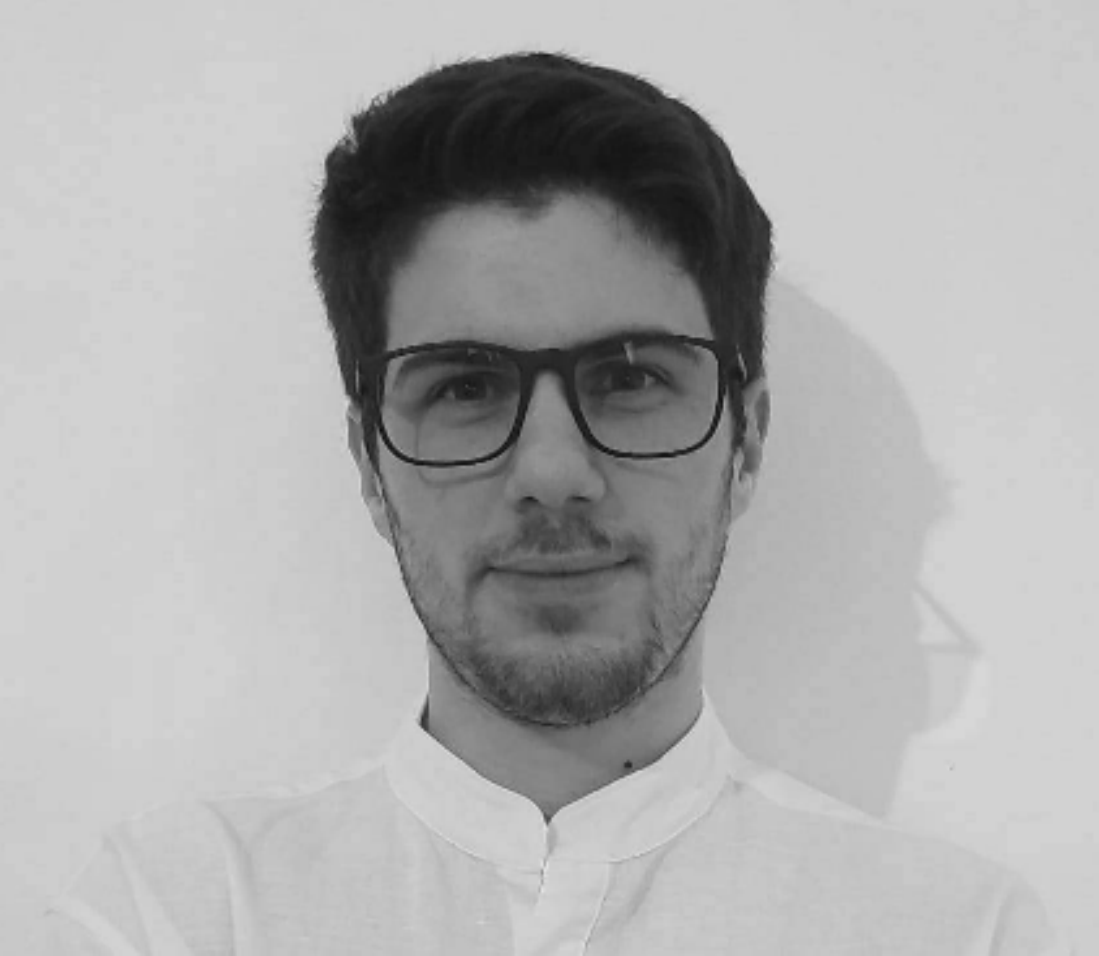}}]{Gabriele Etta}
Gabriele Etta received his MSc degree in Data Science from the University of Padua, Italy, in 2020. He is currently a Ph.D. student from Sapienza University of Rome, Italy, working on Data Driven Modeling of Social Dynamics.
His research interests include complex networks,
information diffusion, and computational social science.
\end{IEEEbiography}

\begin{IEEEbiography}[{\includegraphics[width=1in,height=1.25in,clip,keepaspectratio]{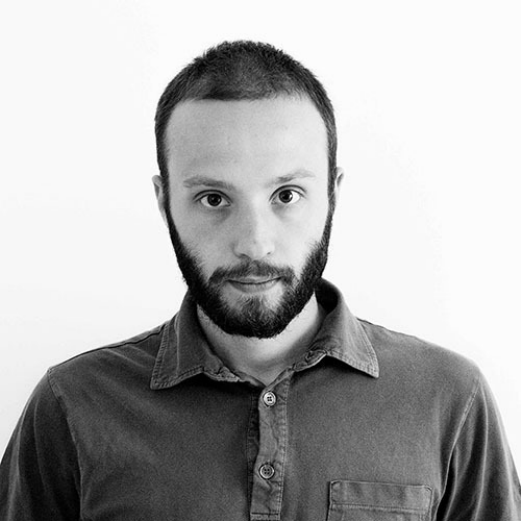}}]{Matteo Cinelli}
Matteo Cinelli is a post-doc researcher at Ca’ Foscari, University of Venice and associate researcher at ISC-CNR. His background is in Management Engineering and he obtained a PhD in Enterprise Engineering from the University of Rome “Tor Vergata”. His research interests include network science, computational social science and big data. 
\end{IEEEbiography}

\begin{IEEEbiography}[{\includegraphics[width=1in,height=1.25in,clip,keepaspectratio]{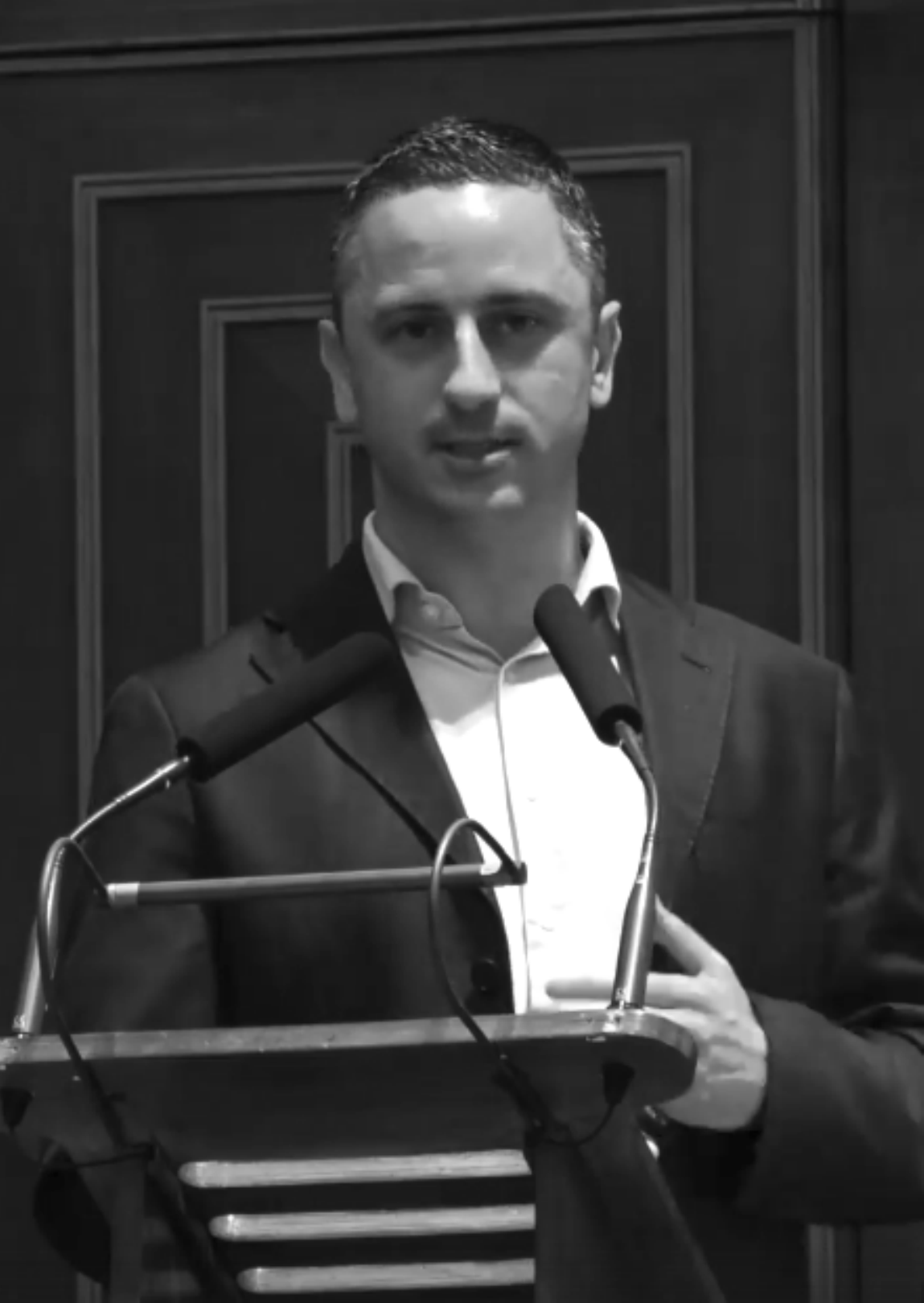}}]{Mauro Conti}

Mauro Conti is Full Professor at the University of Padua, Italy. He is also affiliated with TU Delft and University of Washington, Seattle. He obtained his Ph.D. from Sapienza University of Rome, Italy, in 2009. After his Ph.D., he was a Post-Doc Researcher at Vrije Universiteit Amsterdam, The Netherlands. In 2011 he joined as Assistant Professor the University of Padua, where he became Associate Professor in 2015, and Full Professor in 2018. He has been Visiting Researcher at GMU, UCLA, UCI, TU Darmstadt, UF, and FIU. He has been awarded with a Marie Curie Fellowship (2012) by the European Commission, and with a Fellowship by the German DAAD (2013). His research is also funded by companies, including Cisco, Intel, and Huawei. His main research interest is in the area of Security and Privacy. In this area, he published more than 400 papers in topmost international peer-reviewed journals and conferences. He is Area Editor-in-Chief for IEEE Communications Surveys \& Tutorials, and has been Associate Editor for several journals, including IEEE Communications Surveys \& Tutorials, IEEE Transactions on Dependable and Secure Computing, IEEE Transactions on Information Forensics and Security, and IEEE Transactions on Network and Service Management. He was Program Chair for TRUST 2015, ICISS 2016, WiSec 2017, ACNS 2020, and General Chair for SecureComm 2012, SACMAT 2013, CANS 2021, and ACNS 2022. He is Senior Member of the IEEE and ACM. He is a member of the Blockchain Expert Panel of the Italian Government. He is Fellow of the Young Academy of Europe.
\end{IEEEbiography}

\begin{IEEEbiography}[{\includegraphics[width=1in,height=1.25in,clip,keepaspectratio]{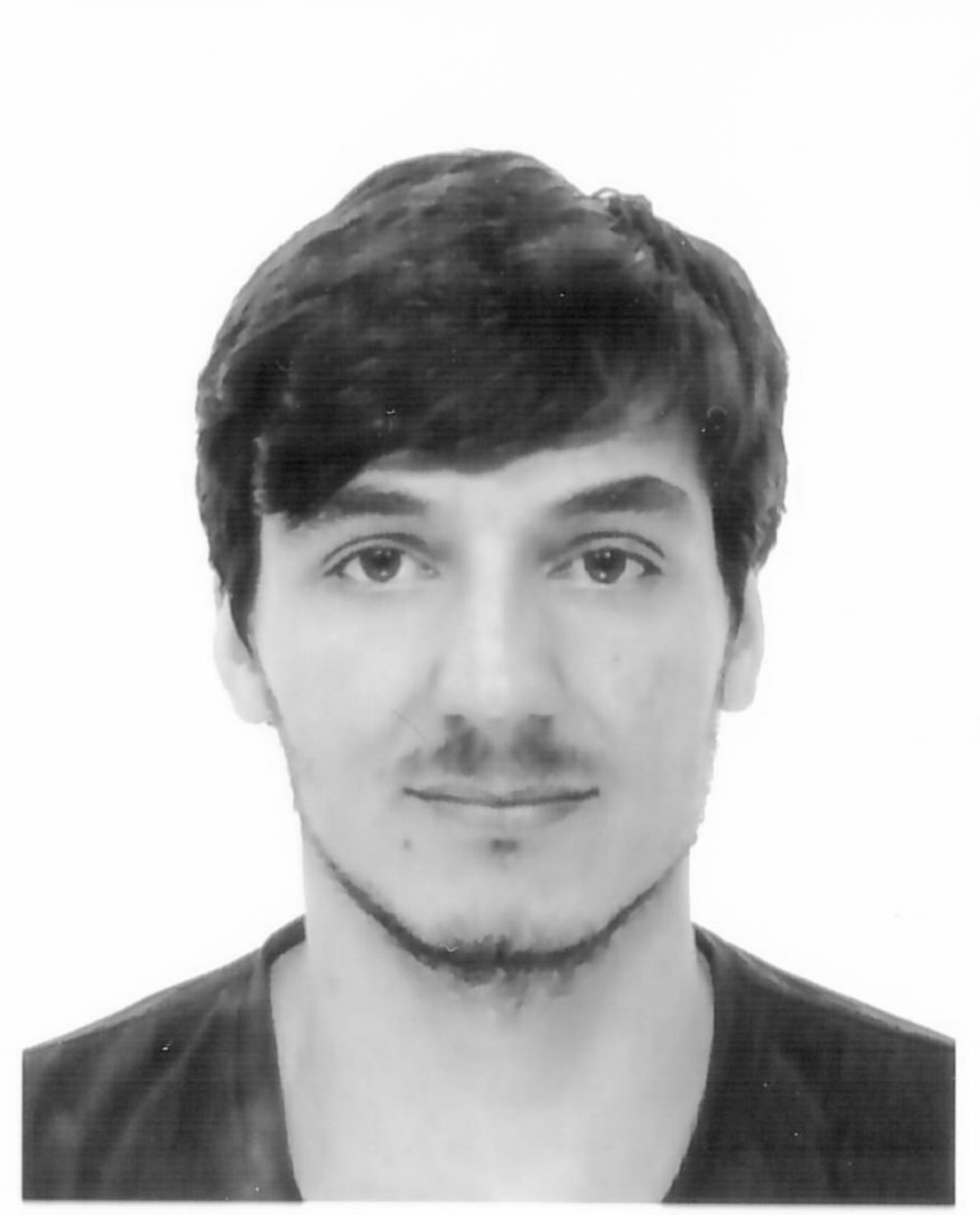}}]{Alessandro Galeazzi}
Alessandro Galeazzi obtained the B.Sc. in Information Engineering in 2016 from the University of Padua. In 2016 he enrolled in the M.Sc. course of ICT for Internet and Multimedia at the University of Padova and in 2017 he joined the double degree program between National Taiwan University and the University of Padova. In 2018 he received the M.Sc. in Communication Engineering from National Taiwan University and the M.Sc. in ICT for Internet and Multimedia from the University of Padova. In 2018 he joined the Ph.D. course at the University of Brescia. His interests include human behavior on online social media, information and misinformation spreading, and social feedback algorithm effects on users’ choices.
\end{IEEEbiography}

\begin{IEEEbiography}[{\includegraphics[width=1in,height=1.25in,clip,keepaspectratio]{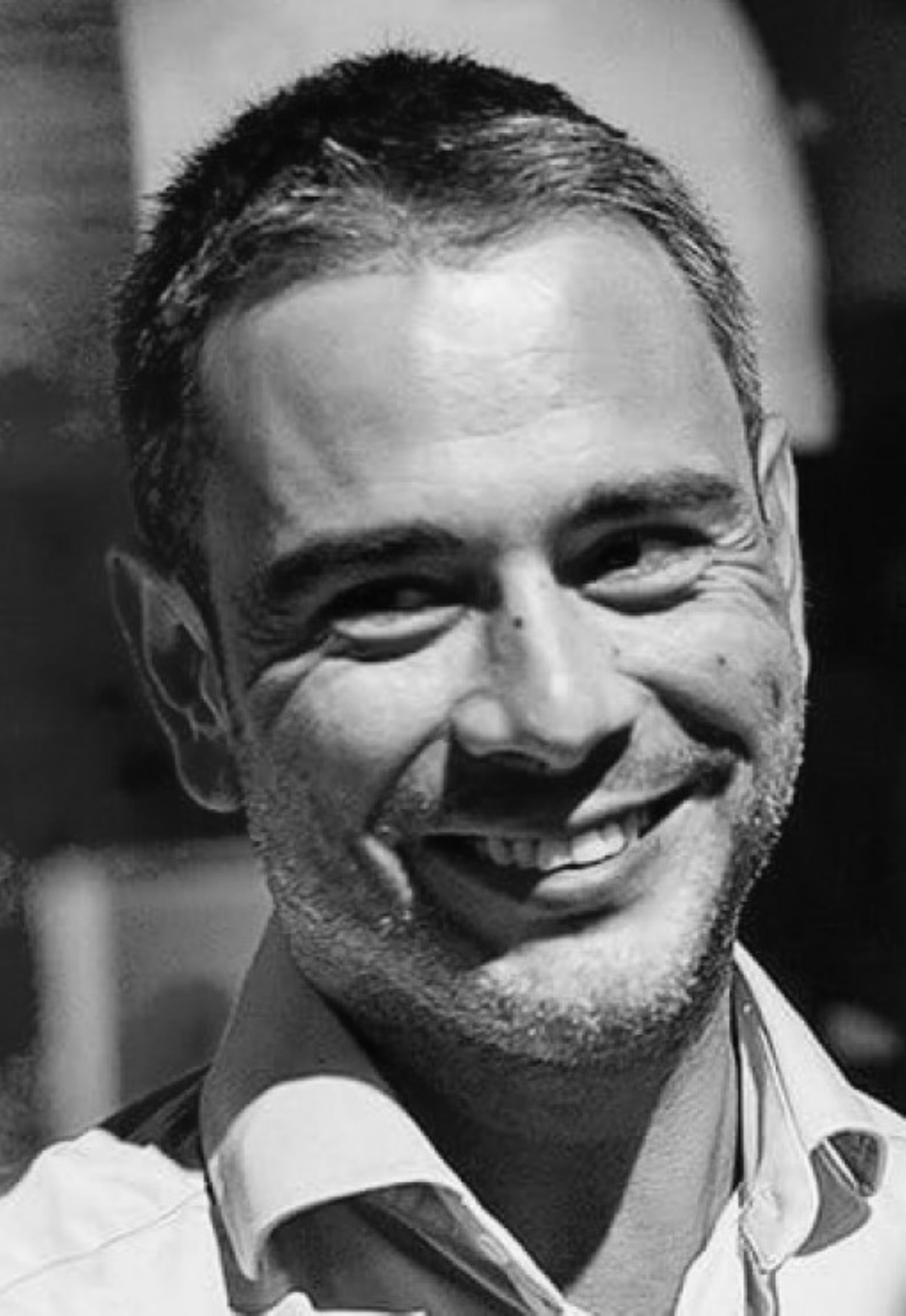}}]{Walter Quattrociocchi}
Walter Quattrociocchi is Associate Professor at Sapienza University of Rome where he leads the Center of Data Science and Complexity for Society (CDCS.
His research interests include data science, network science, cognitive science, and data-driven modeling of dynamic processes in complex networks. His activity focuses on the data-driven modeling of social dynamics such as (mis)information spreading and the emergence of collective phenomena. 
Dr Quattrociocchi has published extensively in peer reviewed conferences and journals including PNAS. The results of his research in misinformation spreading have informed the Global Risk Report 2016 and 2017 of the World Economic Forum and have been covered extensively by international media including Scientific American, New Scientist, The Economist,  The Guardian,  New York Times, Washington Post, Bloomberg, Fortune, Poynter and The Atlantic). He published two books: “Misinformation. Guida alla società dell’informazione e della credulità” (Franco Angeli) and “Liberi di Crederci. Informazione, Internet e Post Verità” with Codice Edizioni for the dissemination of his results.

In 2017 Dr Quattrociocchi was the coordinator of the round table on Fake News and the role of Universities and Research to contrast fake news chaired by the President of Italy's Chamber of Deputies Mrs Laura Boldrini. Since 2018 he is Scientific Advisor of the Italian Communication Authority (AGCOM) and currently Member of the Task Force to contrast Hate Speech nomianted by the Minister of Innovation.
Professor Quattrociocchi is regularly invited for keynote speeches and guest lectures at major academic and other organizations, having presented among others at CERN, European Commission, the University of Cambridge, Network Science Institute, Global Security Forum etc.
\end{IEEEbiography}

\begin{IEEEbiography}[{\includegraphics[width=1in,height=1.25in,clip,keepaspectratio]{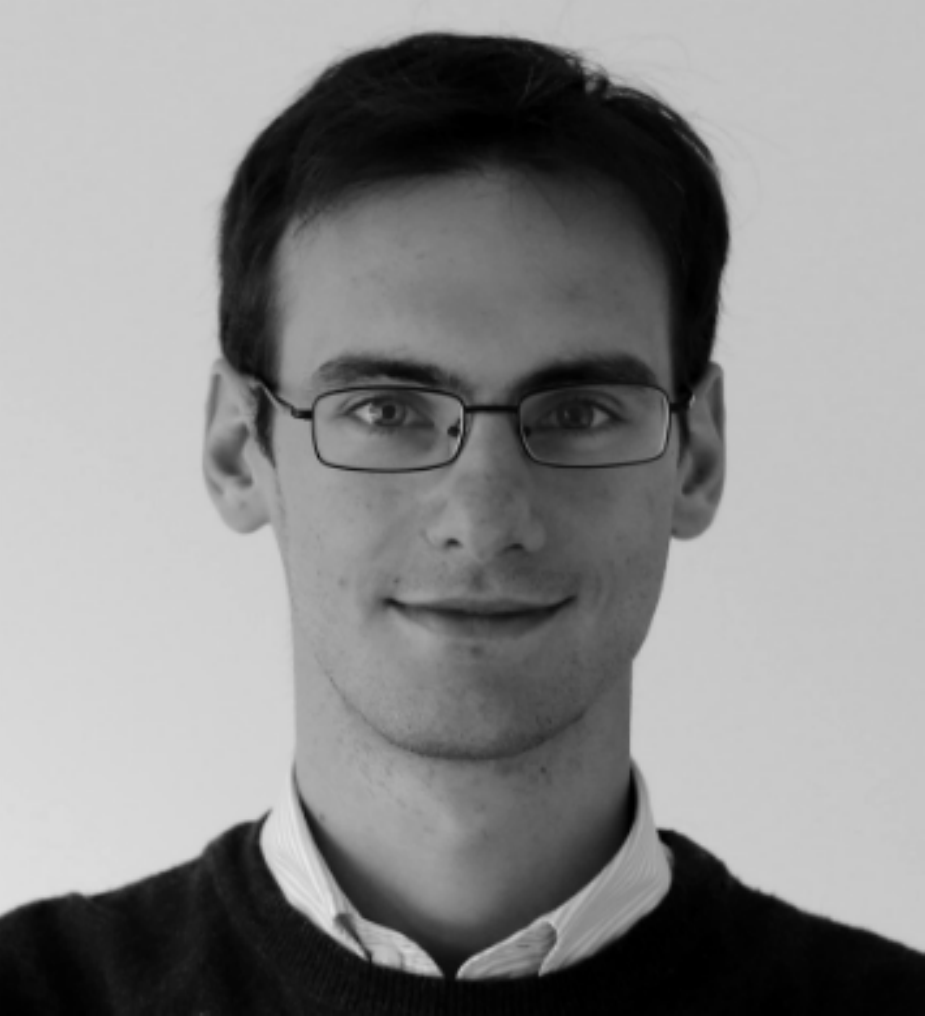}}]{Carlo Michele Valensise}
Carlo Michele Valensise received in 2017 the M.S. degree in Physics from Sapienza University of Rome, and the Ph.D. degree in Physics in 2021, working on ultrafast spectroscopy, and application of Artificial Intelligence to Optics experiments. Besides that, he was Research Fellow of Ca' Foscari University of Venice in Data Science for Complexity and Computational Social Science. He is currently post-doc researcher at Centro Ricerche Enrico Fermi in Rome, working on Computational Social Science.
\end{IEEEbiography}





\end{document}